\documentclass[twocolumn,twoside,slac_two]{revtex4}
\usepackage{graphicx}
\usepackage{fancyhdr}
\usepackage{amsmath}%
\begin{document}

\title{Two- and Three-Body Charmless $\boldsymbol{B}$ Decays at BaBar}

\author{S. Stracka on behalf of the BaBar Collaboration}
\affiliation{Universit\`a degli Studi di Milano and INFN, Sezione di Milano - I-20133 Milano, Italy}
\begin{abstract}
We report recent measurements of rare charmless $B$ decays performed by 
BaBar. The results are based on the final BaBar dataset of 
$424\,{\rm fb}^{-1}$ collected at the 
PEP-II $B$-factory based at the SLAC National Accelerator Laboratory.
\end{abstract}

\maketitle

\thispagestyle{fancy}

%%%%%%%%%%%%%%%%%%%%%%%%%%%%%%%%%%
\section{Introduction}

The study of rare $B$ decays is a key ingredient 
to meet two of the main goals of the $B$-factories: 
assessing the validity of the Cabibbo-Kobayashi-Maskawa (CKM)
picture of $CP$-violation~\cite{CKM} 
by precisely measuring the elements 
of the Unitarity Triangle (UT), and searching for hints of New 
Physics (NP),
or otherwise constraining NP scenarios, in processes 
which are suppressed in the Standard Model (SM). %
In loop processes, in particular, NP at some higher 
energy scale may manifest itself in the low energy effective 
theory as new couplings, such as those introduced by new 
very massive virtual particles in the loop~\cite{MASIERO}. %
In NP searches hadronic uncertainties can play a major role, 
expecially for branching fraction measurements. Many theoretical 
uncertainties cancel in ratios of amplitudes, and most NP probes 
are therefore of this kind.

In the following sections we report recent measurements, 
performed by the BaBar Collaboration, that are relevant to NP searches 
in charmless hadronic $B$ decays. 
In Sec.~\ref{etapks} we report the 2008 update 
of the time-dependent analyses of $B^0\to \eta' K^0$ 
and $B^0 \to \omega K_S$  decays. 
In Sec.~\ref{threebody} we describe the search for 
$B$ decays to the three-body final states $P K_S K_S$, 
where $P=\eta,\eta',\pi^0$. %
The Dalitz Plot analysis of $B\to K_S\pi\pi$ decays 
is reported in Sec.~\ref{dalitz}. 
%This channel can be used to put non-trivial constraints 
%on the $\rho-\eta$ plane (in particular on the CKM 
%angle $\gamma$) and measure $\sin 2\beta_{\rm eff}$.
These measurements are related to the CKM angle $\beta$,
which is experimentally accessible through 
the interference between the direct decay of the $B$ meson 
to a $CP$ eigenstate and the decay after 
$B^0\overline{B}^0$ mixing. This interference 
affects the time evolution of the decay.

In the time-dependent analyses of $B$ decays, one of the 
two $B$ mesons produced in 
$e^+e^-\to \Upsilon(4S) \to B\overline{B}$ events is fully 
reconstructed according to the final state $f$ of interest.
The flavor and the decay vertex position for the other $B$ in 
the event ($B_{tag}$) are identified from its decay products.
The proper time difference between the two $B$ mesons is:
\begin{eqnarray}
\label{decayrate}
\nonumber
f(\Delta t) & = & \frac{e^{-|\Delta t|/\tau}}{4\tau}
\left\{1+ q\left[ -\eta_f S_f\sin(\Delta m_d \Delta t)\right.\right.\\
& & 
\left.\left.-C_f\cos(\Delta m_d \Delta t)
\right]
\right\},
\end{eqnarray}
where $\eta_f$ is the $CP$ eigenvalue of the final state $f$, 
$q=+1(-1)$ if the $B_{tag}$ decays as a $B^0$ ($\overline{B}^0$), 
$\tau=(1.536\pm 0.014)\,{\rm ps}$~\cite{PDG} is the mean 
$B$ lifetime, and 
$\Delta m_d=(0.502\pm 0.007)\,{\rm ps}^{-1}$ is the 
$B^0-\overline{B}^0$ mixing frequency~\cite{PDG}. 

The angle $\beta$ can be determined very accurately and 
with small theoretical uncertainties from the analysis of the 
decay rate asymmetries in $b\to c\bar{c}s$ decays, 
and is related in a simple way to the parameter 
$S_{cc}$ in Eq.~\ref{decayrate}:
\begin{equation}
S_{cc}=-\eta_f\sin 2\beta.
\end{equation}
The current world average is $S_{cc}=0.67\pm 0.02$~\cite{HFAG}.

A parameter $S_{qq}=-\eta_{CP}\sin 2\beta_{\rm eff}$ 
(where $\beta_{\rm eff}$ denotes an effective value of $\beta$) 
can also be extracted from penguin (loop) dominated 
$b\to q\bar{q}s$ decays. These decays may receive  
contributions from NP effects that can lead to 
measurable differences $\Delta S\equiv S_{qq} - S_{cc}$. 
In some NP scenario, deviations can be $\approx O(1)$~\cite{NPORDERONE}.
A background to $\Delta S$ measurements 
as a tool for NP searches 
is represented by the CKM suppressed tree contributions, which may
introduce shifts in the parameter $S$. % 
The SM effect is predicted in most models to be a positive 
shift in $\Delta S$~\cite{SMPOSSHIFT}.
The size of this shift is related to the ratio of tree 
to penguin amplitudes, which depends on the decay mode.
Theoretical estimates for this ratio are affected by 
large uncertainties for most decay modes, 
with the exception of the $K_S K_S K^0$, $\phi K^0$, 
%modes (in which tree amplitudes don't contribute), 
and $\eta' K^0$ modes. %(in which the gluon penguin amplitude is enhanced).

The updated measurement of 
the direct $CP$-violation parameter for $B\to K^+\pi^-$ 
is reported in Sec.~\ref{kpipuzzle}.
$B\to K\pi$ transitions receive contributions 
from tree, color suppressed tree, gluonic penguin (loop), 
and electroweak penguin contributions.
Among these amplitudes, the dominant contribution is 
represented by the gluonic penguin.
The electroweak penguin and the color 
suppressed tree amplitudes are expected to be small in the SM, 
and the direct $CP$-violation parameter for $K^+\pi^-$
decays should be equal to that for $K^+\pi^0$ decays.
The rates and direct $CP$-violating 
asymmetries in the $K\pi$ system reveal puzzling 
features, that might be indications of NP.

%%%%%%%%%%%%%%%%%%%%%%%%%%%%%%%%%%
\section{$\boldsymbol{B^0 \to \eta' K^0}$ and $\boldsymbol{B^0\to \omega K_S}$}\label{etapks}
The BaBar Collaboration has updated the measurement of $CP$-violating 
asymmetries in $B$ decays to $\eta' K^0$ and $\omega K_S$ using the 
final dataset corresponding to  $465$ million $B\overline{B}$ 
pairs~\cite{ETAPKX}.

\begin{figure}[h]
\centering
\includegraphics[height=90mm]{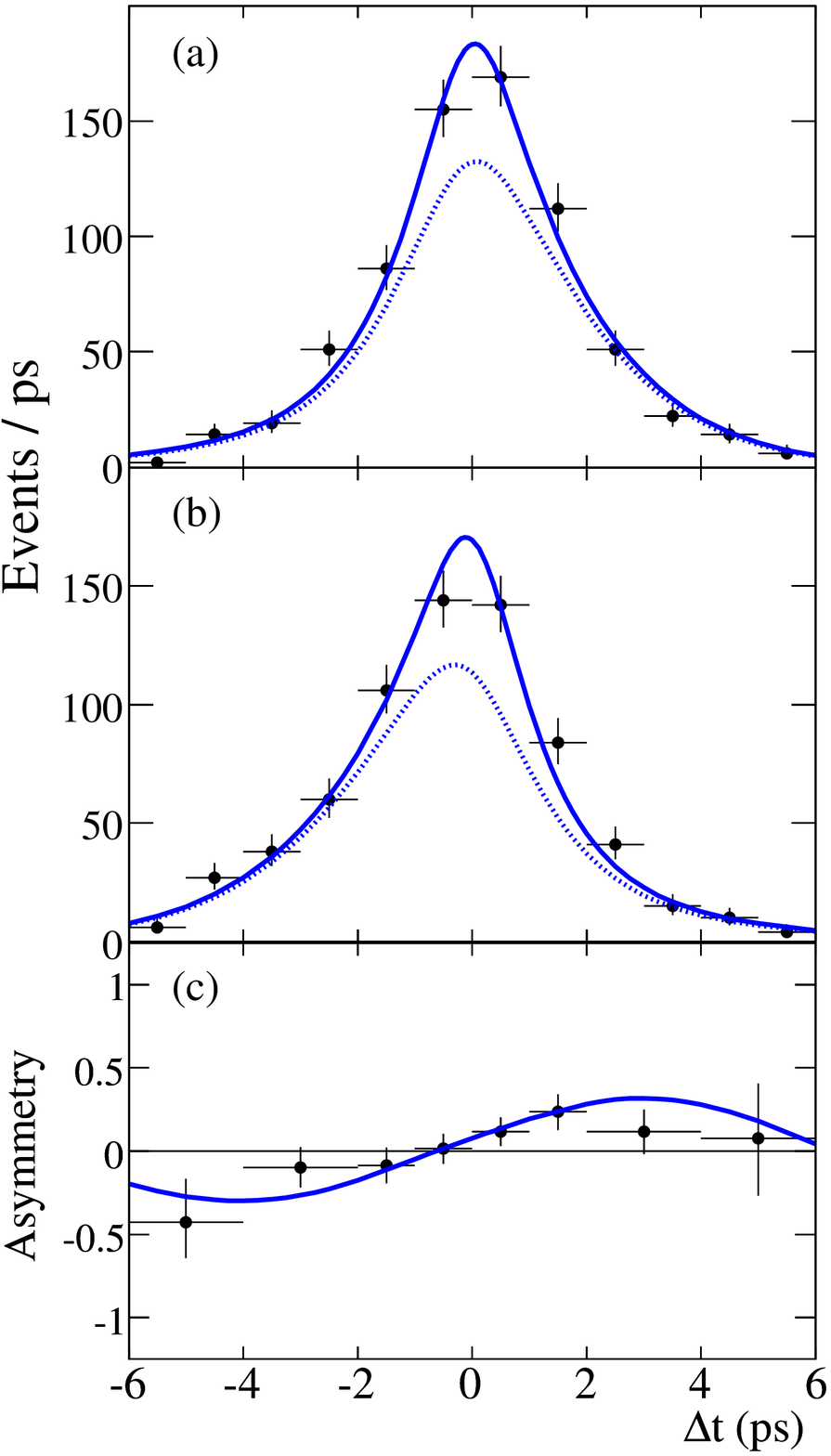}\\
\includegraphics[height=90mm]{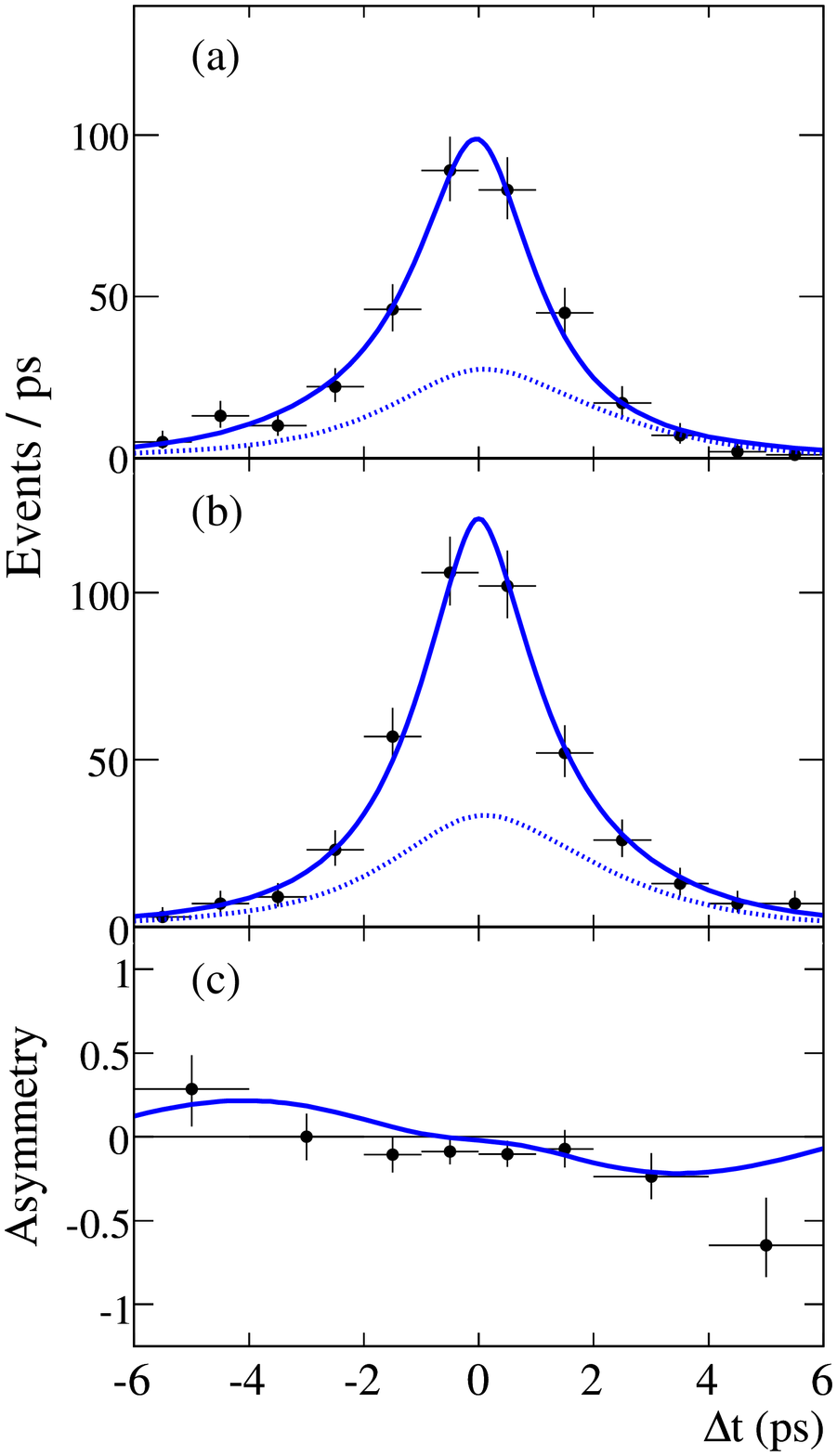}
\caption{Data and model projections for $\eta' K_S$ (top) and 
$\eta' K_L$ (bottom) onto $\Delta t$ 
for (a) $B^0$ and (b) $\overline{B}^0$ tags. Points with error 
bars represent the data: the solid (dotted) line displays the total
fit function (total signal). In (c) we show the raw asymmetry, 
$(N_{B^0}-N_{\overline{B}^0})/(N_{B^0}+N_{\overline{B}^0})$:
the solid line represents the fit function~\cite{ETAPKX}.} 
\label{etapdt}
\end{figure}

The $\eta'K^0$ channel provides the most precise measurement 
of $S$ in a penguin dominated mode. 
The time-dependent analysis of the $B\to \eta' K^0$ 
decays is performed on about $2500$ signal events, 
corresponding to a branching fraction of 
$\approx 65 \times 10^{-6}$. The resulting decay rate 
distributions are shown in Fig.~\ref{etapdt}.
Both $\eta'K_S$ and $\eta' K_L$ modes are analyzed, 
and the results are combined in scans of  $-\ln{\cal L}$, 
taking into account the different intrinsic $CP$ parity. 
With respect to the previous measurement, the analysis relies 
on about $20\%$ more data, improved track reconstruction, and 
reconstruction of an additional $\eta'$ decay channel in $\eta' K_L$ analysis.
The results for the $CP$-violation parameters $S$ and $C$ 
in this decay mode are~\cite{ETAPKX}: 
\begin{eqnarray}
S_{\eta'K^0} & = & \phantom{-}0.57\pm 0.08 \pm 0.02\\
C_{\eta'K^0} & = & -0.08\pm 0.06 \pm 0.02,
\end{eqnarray}
where the first error is statistical and the second systematic.
%The main contribution to systematic uncertainty comes 
%from the vertex resolution model.
The uncertainties on $S$ and $C$ have decreased by about $20\%$ and $25\%$, 
respectively, and are still statistically limited.
The observed discrepancy between $S_{cc}$ and $S_{\eta'K^0}$
is not significant. 
Since this is one of the theoretically cleanest modes, 
the study of decay rate asymmetries in 
$B^0\to \eta'K^0$ decays is a benchmark analysis for a 
Super $B$ factory~\cite{SUPERB}.

In the $B^0\to \omega K_S$ channel 
the tree to penguin ratio is not necessarily small.
If the two contributions are comparable in size they can lead 
to a non-zero direct $CP$ parameter. 
For the $\omega K_S$ channel, $B$ daughters are reconstructed in 
the main decay modes, 
$\omega \to \pi^+\pi^-\pi^0$ and $K_S\to \pi^+\pi^-$.
The $\omega$ mass and angular variables are included in the fit.
The signal yield for the $\omega K_S$ channel is much lower ($N=163$) 
than that for the $\eta' K^0$ channel, 
and the measurement of $S$ and $C$ is therefore less precise. The extracted 
values for these parameters are~\cite{ETAPKX}:
\begin{eqnarray}
S_{\omega K_S}& = & \phantom{-}0.55^{+0.26}_{-0.29} \pm 0.02\\
C_{\omega K_S}& = & -0.52^{+0.22}_{-0.20} \pm 0.03
\end{eqnarray}
%Here, the main systematic uncertainty arises from PDF characterization.
The central value of $C$ is of potential interest as it 
differs from $0$ by more than $2\sigma$, but 
more statistics than that collected at the $B$-factories 
is needed to further investigate direct $CP$-violation in this decay.

\clearpage

%%%%%%%%%%%%%%%%%%%%%%%%%%%%%%%%%%
\section{$\boldsymbol{B^0 \to X^0 X^0 P^0}$}
\label{threebody}
Decays to $CP$-eigenstate final states with three $CP$ eigenstate particles, 
two of which are equal, allow the measurement of $\beta_{\rm eff}$ from 
time-dependent $CP$-asymmetries.
Since the $CP$-eigenvalue is independent of the 
intermediate resonant structure, there is no need for an isospin 
or Dalitz Plot analysis to separate the different resonant 
contributions~\cite{GERSH}.
BaBar has searched for these decay channels in 
$426\,{\rm fb}^{-1}$ data~\cite{PPX}.
A maximum likelihood fit to kinematical and topological 
variables is performed but no evidence of signal 
is found (see Table~\ref{resppx}), 
and yields are too small to allow a time-dependent analysis.

\begin{table}[h]
\begin{center}
\caption{Branching fraction results and upper limits at 90\% confidence level (CL) for $B^0 \to X^0 X^0 P^0$ channels~\cite{PPX}.}
\begin{tabular}{|l|c|c|}
\hline \textbf{Mode} & $\mathbf{{\cal B} (\times 10^{-6}})$ & \textbf{90\% CL UL} $\mathbf{ (\times 10^{-6})}$ \\
\hline $\pi^0K_SK_S$ & $2.7^{+4.2}_{-3.7}\pm 0.6$ & $9$ \\
\hline $\eta K_SK_S$ & $2.7^{+4.7}_{-3.8}\pm 1.2$ & $10$ \\
\hline $\eta' K_SK_S$ & $2.7^{+8.0}_{-6.5}\pm 3.4$ & $20$ \\
\hline
\end{tabular}
\label{resppx}
\end{center}
\end{table}

%%%%%%%%%%%%%%%%%%%%%%%%%%%%%%%%%%
\section{$\boldsymbol{B^0 \to K_S\pi^+\pi^-}$}
\label{dalitz}
The $S$ parameter can be extracted 
from a time-dependent Dalitz Plot analysis of $B\to K_S\pi\pi$ 
transitions.
BaBar has performed an analysis of this decay using a 
data sample of $383$ million $B\overline{B}$ pairs~\cite{DALITZ}. 

The decay amplitude can be parameterized as a function of two Mandelstam-like
variables 
$s_+\equiv m^2_{K^0_S\pi^+}$ and  
$s_-\equiv m^2_{K^0_S\pi^-}$ 
according to the isobar model:
\begin{eqnarray}
A(s_+, s_-)&=&\sum^{N}_{j=1}|c_j|e^{-i\phi_j} \times \\ \nonumber
& & R_j(m)X_L(|\vec{p}^*|r')X_L(|\vec{q}|r)T_j(L,\vec{p},\vec{q}).
\end{eqnarray}
Here the complex coefficients $|c_j|$ and $e^{-i\phi_j}$ are the 
relative magnitude and phase for the channel $j$, 
$R_j(m)$ is the lineshape term, 
$X_L$ are Blatt-Weisskopf barrier factors~\cite{BLATTWEIS}, 
$T_j$ is the angular distribution,
$\vec{p}$ is the momentum of the bachelor particle, 
$\vec{q}$ is the momentum of one of the resonance daughters, $L$ is
the orbital angular momentum between $\vec{p}$ and the resonance 
momentum, and a star denotes the $B$-meson rest frame.
The signal model includes both $\pi\pi$ ($\rho(770), f_0(980), f_2(1270), 
f_X(1300),\chi_{c0}$) and $K\pi$ ($K^*(892), (K\pi)_{S-wave}$) resonances
as well as non-resonant contributions, for a total of 15 complex 
scalar amplitudes.

\begin{figure}[t]
\centering
\includegraphics[width=80mm]{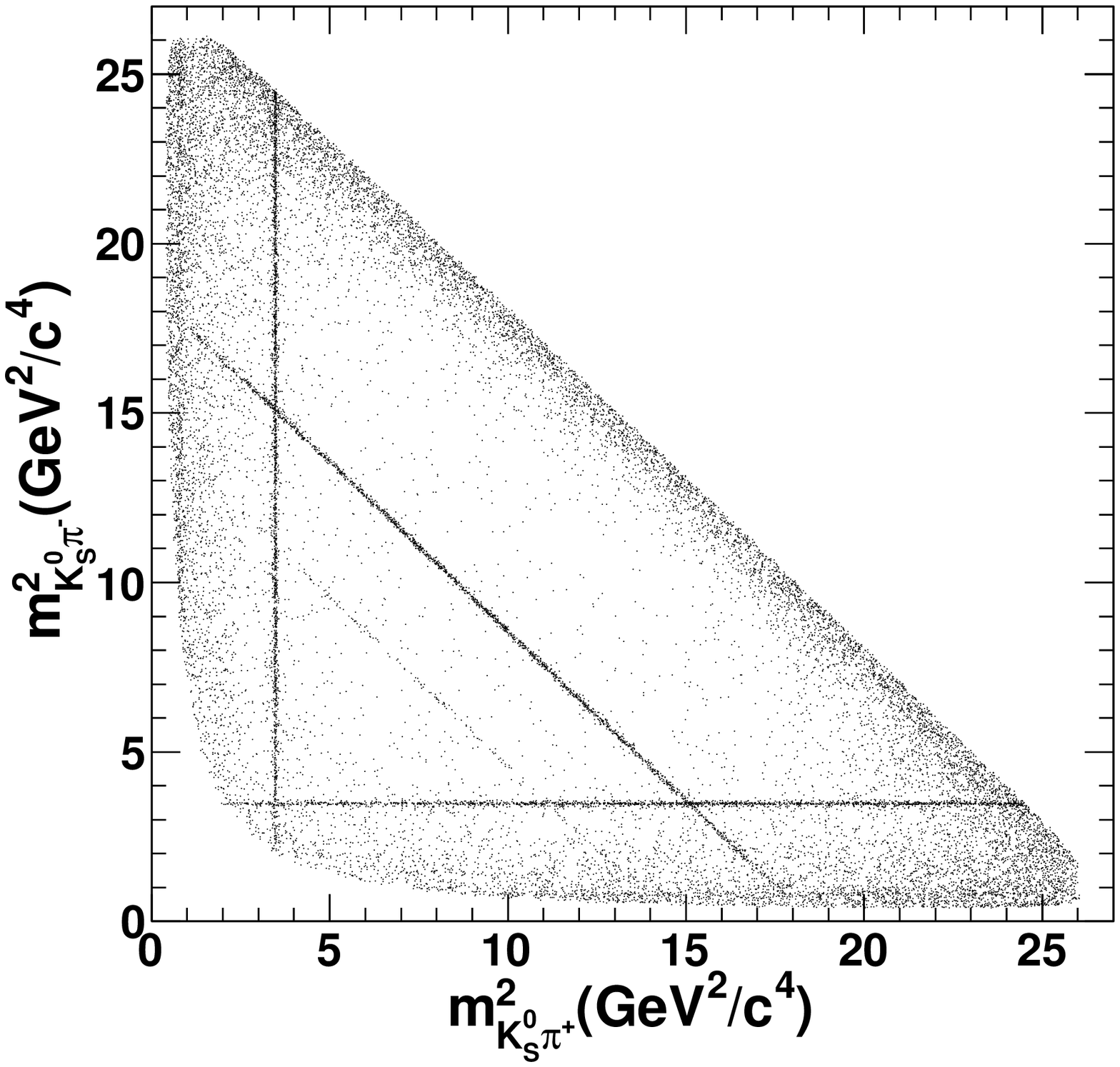}\\
\includegraphics[width=80mm]{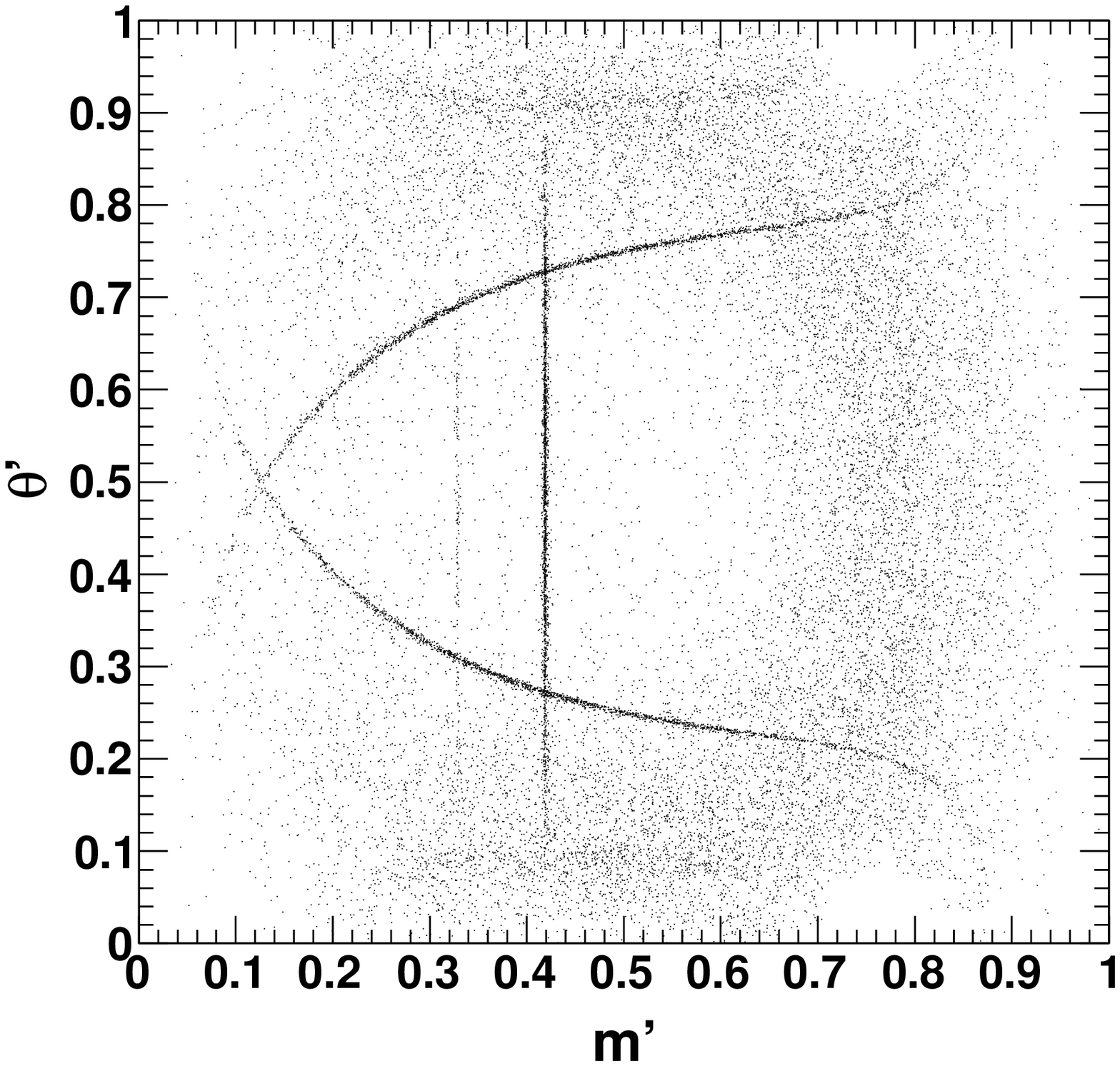}
\caption{Standard (top) and square (bottom) Dalitz plots
of $B^0 \to K_S \pi^+ \pi^-$ candidates~\cite{DALITZ}.} \label{sdp}
\end{figure}

The choice of $s_+$ and $s_-$ as independent variables is impractical 
because both the signal events and the combinatorial $e^+ e^- \to q\overline{q}$
events populate the kinematic boundaries of the Dalitz Plot. 
The following new coordinates are therefore introduced: 
\begin{eqnarray}
  m'\equiv \frac{1}{\pi}\arccos\left( 
  2\frac{m_0-m_0^{\rm min}}{m_0^{\rm max}-m_0^{\rm min}} -1 
  \right), & & \theta'\equiv \frac{1}{\pi} \theta_0, \nonumber
\end{eqnarray}
where $m_0$ is the $\pi^+\pi^-$ invariant mass, 
$m_0^{\rm max}$ and $m_0^{\rm min}$ are the kinematic limits of $m_0$, and 
$\theta_0$ is the $\pi^+\pi^-$ resonance helicity angle.
The following transformation is applied, 
\begin{equation}
ds_+ds_- \to |\det J| dm' d\theta'
\end{equation}
which defines the square Dalitz Plot shown in Fig.~\ref{sdp}.

\begin{figure*}[t]
\centering
\includegraphics[height=100mm]{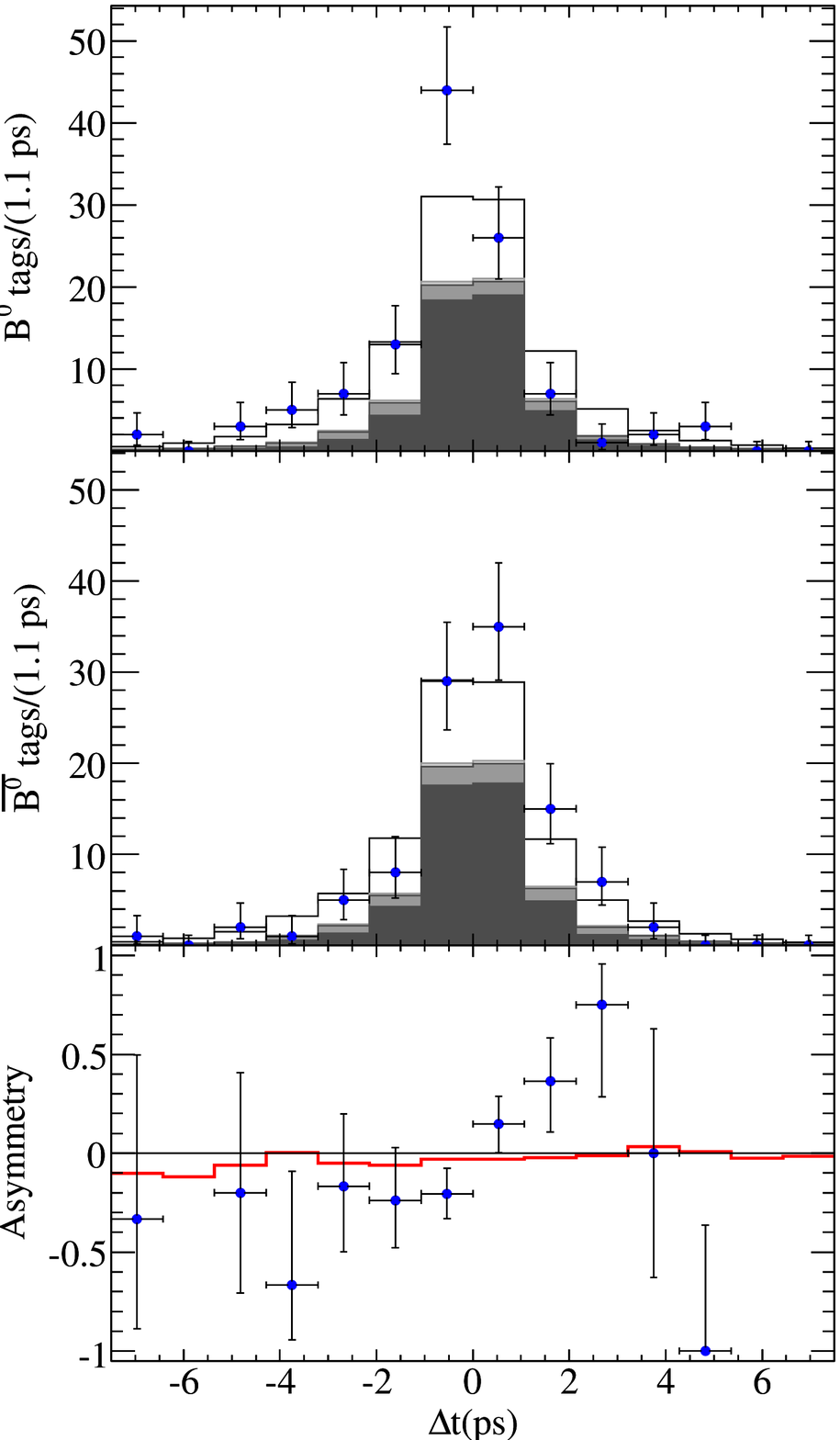}
\includegraphics[height=100mm]{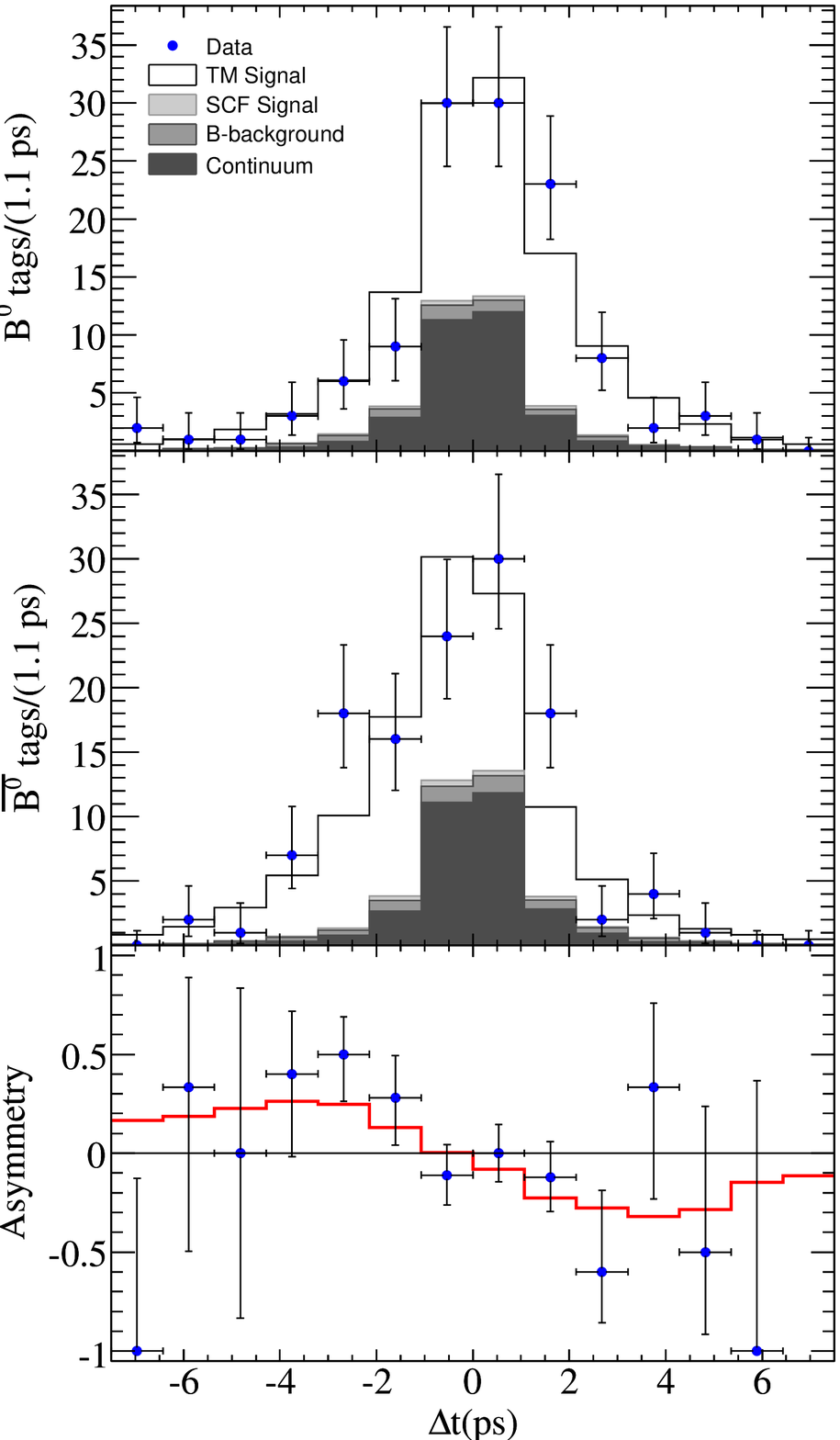}
\caption{Distributions of $\Delta t$ when the $B_{tag}$ is a $B^0$ (top), $\overline{B}^0$ (middle), and the derived $\Delta t$ asymmetry (bottom). 
Plots on the left (right) hand side, correspond to events in the
$f_0(980) K_S$ ($\rho(770)^0 K_S$) region~\cite{DALITZ}.} \label{deltat}
\end{figure*}

The distribution of the proper-time difference between the two 
$B$-mesons is given by:
\begin{eqnarray}
\nonumber
f(\Delta t) & = & \frac{e^{-|\Delta t|/\tau}}{\tau} \left[ 
  |A|^2 + |\overline{A}|^2 \right. \\ 
\nonumber
& & \mp \left(|A|^2 - |\overline{A}|^2\right) \cos(\Delta m_d \Delta t) \\
\nonumber
& & \left. 
\pm \eta_f 2{\rm Im}\left[\overline{A} A^*\right]\sin(\Delta m_d \Delta t) 
\right],
\end{eqnarray}
where the $CP$-violation parameters are expressed in terms of the 
amplitudes $A$ (for $B$ decays) and $\overline{A}$ (for $\overline{B}$ decays).
From the results of this analysis for $B^0$ decays to 
$\rho(770)^0K_S$ and $f_0(980)K_S$ $CP$-eigenstates 
it is possible to calculate $\beta_{\rm eff}$ as: 
\begin{equation}
\beta_{\rm eff} = \frac{1}{2}\arg\left(c_k\overline{c}_k^*\right).
\end{equation}
The direct and mixing-induced $CP$ asymmetries can be calculated as:
\begin{eqnarray}
C_k&=&\frac{|c_k|^2-|\overline{c}_k|^2}{|c_k|^2+|\overline{c}_k|^2}\\
S_k&=&\frac{2{\rm Im} \left(c_k^*\overline{c}_k\right)}{|c_k|^2+|\overline{c}_k|^2}.
\end{eqnarray}
%The interference between different resonant contributions can resolve 
%the ambiguity on $\beta_{\rm eff}$.
Signal enriched distributions for $\Delta t$ and $\Delta t$-asymmetry 
for events in the regions of $\rho(770)$ and $f_0(980)$ are 
shown in Fig.~\ref{deltat}.
\begin{figure*}[t]
\centering
\includegraphics[height=80mm]{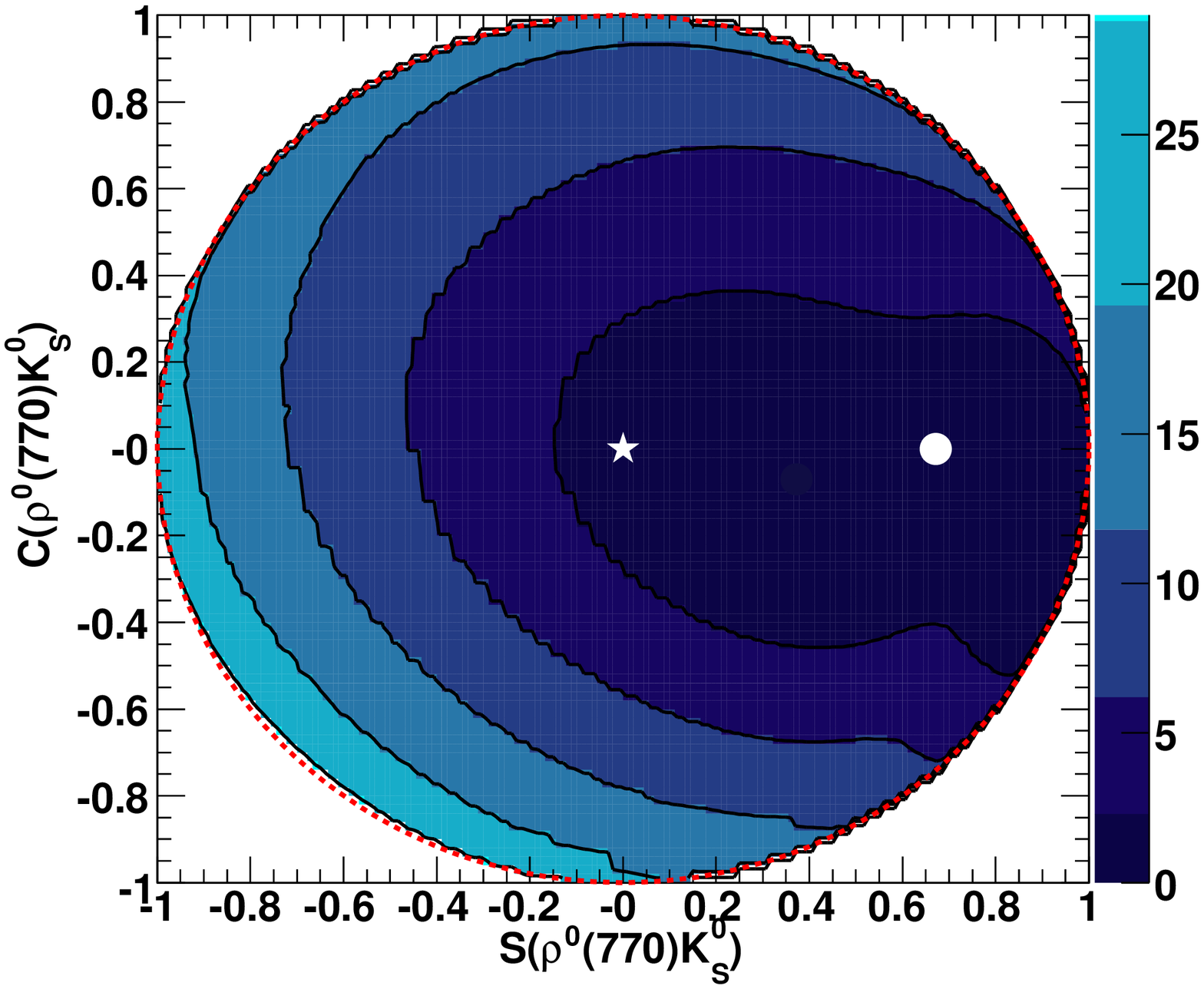}
\includegraphics[height=80mm]{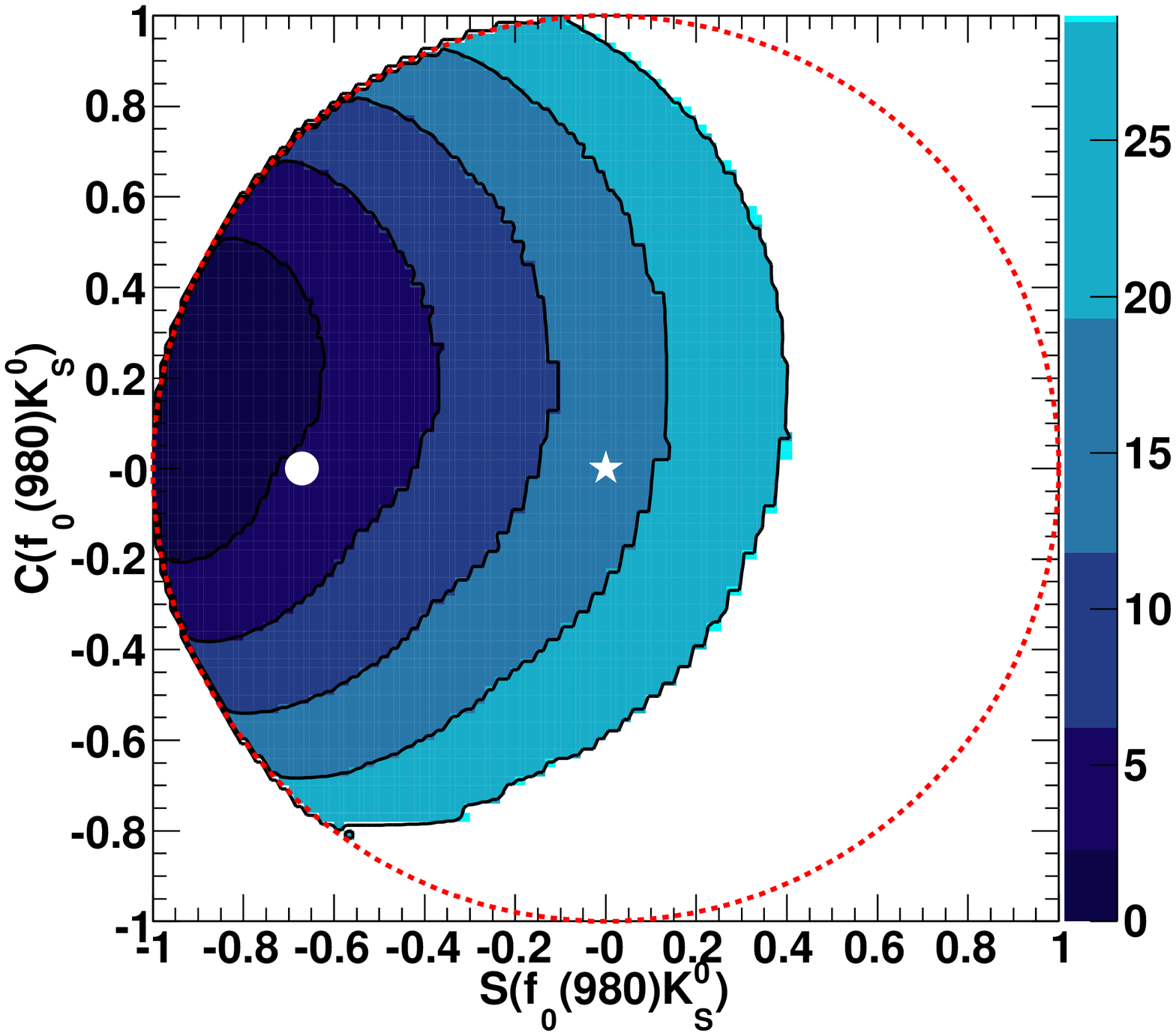}
\caption{Two-dimensional scans of $-2\Delta \log {\cal L}$ as a function of 
($S$, $C$), for the $f_0(980)K_S$ (left) and $\rho(770)^0 K_S$ (right) isobar
components~\cite{DALITZ}. 
The value $-2\Delta \log {\cal L}$ is computed including 
systematic uncertainties.
Shaded areas, from the darkest to the lightest, represent the one to five 
standard deviations contours. The $\bullet$ ($\star$) marks the expectation 
ased on the 
current world average from $b \to c\overline{c}s$ modes~\cite{HFAG} 
(zero point). The dashed circle represents the physical border 
$S^2 + C^2 = 1$.} \label{q2bSC}
\end{figure*}
\begin{figure*}[t]
\centering
\includegraphics[width=80mm]{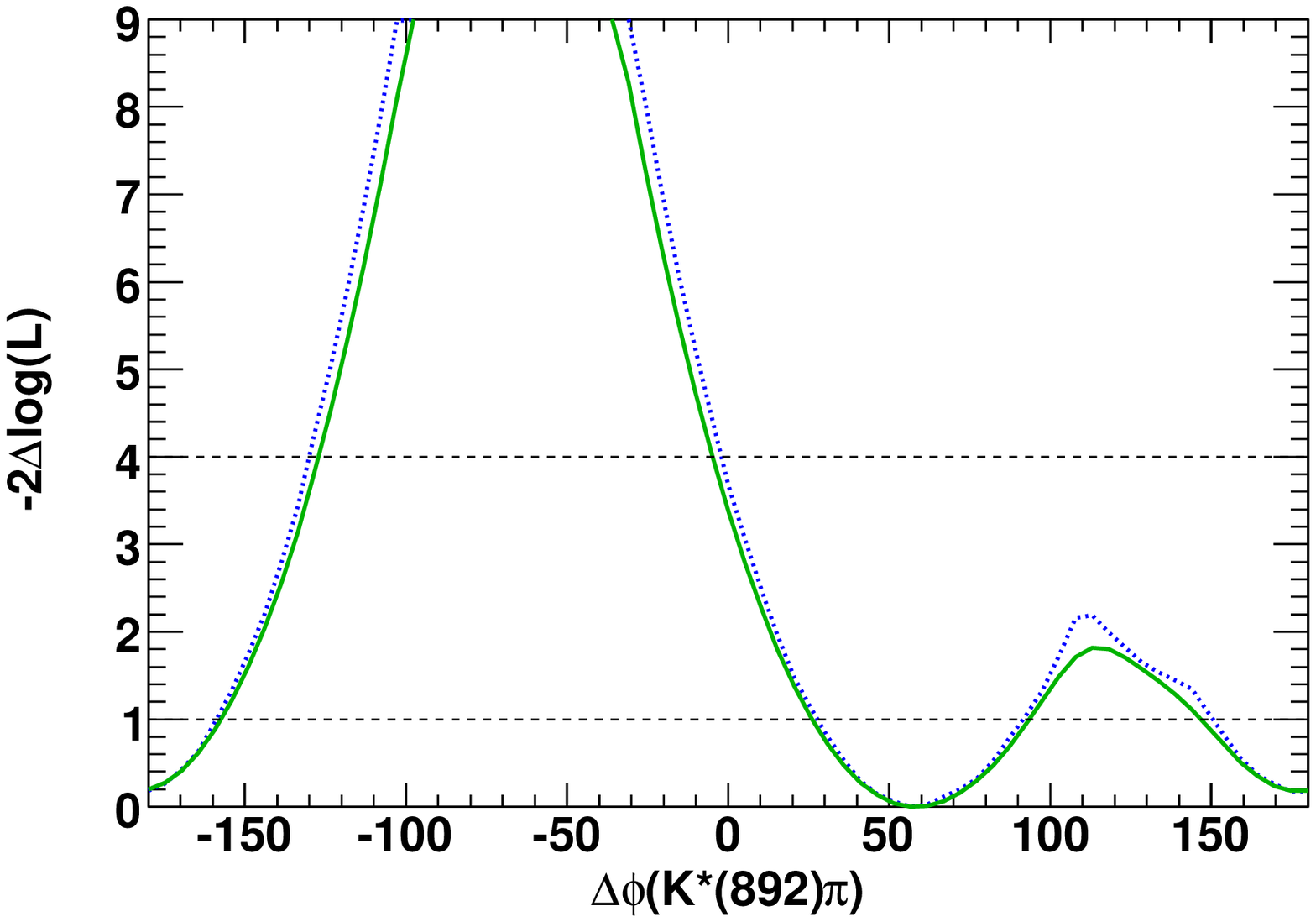}
\includegraphics[width=80mm]{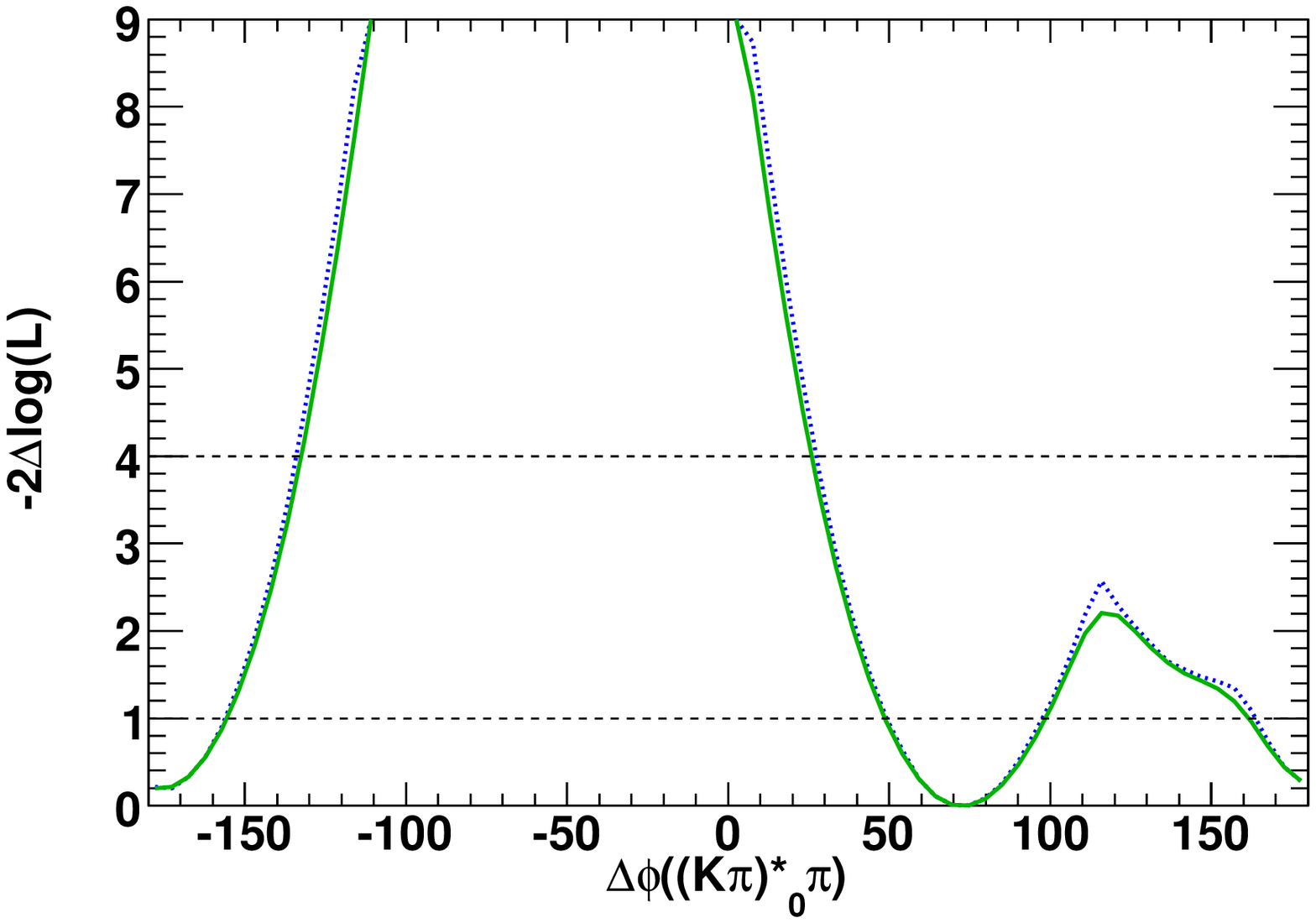}
\caption{Statistical (dashed line) and total (solid line) scans of 
$-2\Delta \log {\cal L}$ as a function of the relative phases 
$\Delta \Phi(K^*(892)\pi)$ (left) and $\Delta \Phi((K\pi)^*_0)$ 
(right)~\cite{DALITZ}. 
Horizontal dotted lines mark the one and two standard deviation levels.} 
\label{scanphdiff}
\end{figure*}
Two solutions for the $S$ and $C$ parameters are present, 
almost degenerate in likelihood~\cite{DALITZ}.
For the $f_0(980)K_S$ channel, there is an evidence of non-zero $S$ at 
about $3.5\sigma$ significance (see Fig.~\ref{q2bSC}). 
Including systematic and Dalitz plot model uncertainties, the 
confidence intervals at $95\%$ confidence level (CL) for the measured value 
of $\beta_{\rm eff}$ in the $B^0 \to \rho^0(770)K_S$ and 
$B^0 \to f_0(980)K_S$ channels are~\cite{DALITZ}:   
\begin{eqnarray}
-9^{\circ}<&\beta_{\rm eff}(\rho^0(770)K_S)&<57^{\circ},\\
18^{\circ}<&\beta_{\rm eff}(f_0(980)K_S)&<76^{\circ}
\end{eqnarray}
These measurements are consistent with the SM.

The analysis of $B^0\to K_S\pi^+\pi^-$ decays provides further non-trivial 
constraints on the $\rho-\eta$ plane (CKM angle $\gamma$). 
In particular, the phase 
difference between isobar amplitudes for $B^0$ and $\overline{B}^0$
decays to $K^*(892)^+\pi^-$ and $(K\pi)_{S-wave}^+\pi^-$ 
\begin{eqnarray}
\nonumber \Delta \phi_{K^*\pi} & \equiv & \arg c_{K^*(892)^+\pi^-} - \arg c_{K^*(892)^-\pi^+}\\
\nonumber \Delta \phi_{(K\pi)_S\pi} & \equiv & \arg c_{(K\pi)_S^+\pi^-} - \arg c_{(K\pi)_S^-\pi^+}
\end{eqnarray}
can be used to extract information about $\gamma$~\cite{CIUCHINI}.
The one-dimensional scans for $\Delta \phi_{K^*\pi}$ and 
$\Delta \phi_{(K\pi)_S\pi}$ are presented in Fig.~\ref{scanphdiff}. Due to 
the small overlap between the phase space regions accessible to $B^0$ and 
$\overline{B}^0$, the sensitivity to these phases is limited.
The available statistics allows to exclude only small regions at $95\%$ CL.
Including systematic effects, the excluded regions are~\cite{DALITZ}:   
\begin{eqnarray}
\nonumber -137^{\circ} < & \Delta\phi_{K^*(892)\pi}& < -5^{\circ}, \\
\nonumber -132^{\circ} < & \Delta\phi_{(K\pi)_S\pi}& < +25^{\circ}.
\end{eqnarray}

%

%%%%%%%%%%%%%%%%%%%%%%%%%%%%%%%%%%
\section{$\boldsymbol{B^0 \to K^+\pi^-}$}
\label{kpipuzzle}

The direct $CP$ violation $A_{CP}$ parameter for $B\to K^+\pi^-$ has 
been extracted from a fit to the $\pi^+\pi^-$, $K^{\pm}\pi^{\mp}$, and $K^+K^-$ 
final states, using $467$ million $B\overline{B}$ pairs~\cite{BTOPIPI}.
Since $K^+\pi^-$ is a self-tagging mode, $A_{CP}$ is obtained 
by simple event-counting:
\begin{equation}
A_{CP}=\frac{N(\overline{B}^0\to K^-\pi^+)-N(\overline{B}^0\to K^+\pi^-)}
{N(\overline{B}^0\to K^-\pi^+)+N(\overline{B}^0\to K^+\pi^-)}.
\end{equation}
This analysis yields a $6.1\sigma$ observation of direct $CP$ violation, 
$A_{CP}(K^+\pi^-)=-0.107\pm 0.016^{+0.006}_{-0.004}$~\cite{BTOPIPI}.
By combining this result with $A_{CP}(K^+\pi^0)$ and taking 
the averages with Belle~\cite{HFAG}, one obtains:  
\begin{eqnarray}
A_{CP}(K^+\pi^-)& = & -0.098^{+0.012}_{-0.011}\\
A_{CP}(K^+\pi^0)& = & +0.050\pm 0.025.
\end{eqnarray}
This result corresponds to a $5\sigma$ deviation from 
the SM expectation. 

Several interpretations have been proposed for this 
discrepancy. In some models this result 
is considered as a hint of NP, which might manifest
itself with large contributions from electroweak 
penguins~\cite{EWP}.
Standard Model explanations have also been formulated, 
which involve large color-suppressed trees and nonperturbative
effects~\cite{CST}.

\section{Conclusions}
In this paper we have reported recent results 
on charmless hadronic $B$ decays obtained by the BaBar experiment.
These results can be used to set non-trivial constraints on the 
Unitarity Triangle, most notably on $\sin 2\beta$ and $\gamma$.

Penguin dominated decays provide several observables 
that can be used as NP probes. 
No hint of NP has been found in the channels 
described in this report, 
most deviations being within $2\sigma$ 
from Standard Model predictions. 
For the theoretically cleanest modes, the experimental uncertainty
on $\Delta S$ measurements is dominant, thus demanding for more statistics 
and precision.

In $B\to K\pi$ decays a $5\sigma$ discrepancy with respect to naive 
Standard Model expectations is observed. The interpretations 
range from SM explanations to NP effects.

%%%%%%%%%%%%%%%%%%%%%%%%%%%%%%%%%%
\begin{acknowledgments}
  I would like to thank the organizers of DPF 2009 for an interesting 
  conference and my BaBar and PEP-II collaborators for their contributions.
\end{acknowledgments}

\bigskip % extra skip inserted


\end{document}